\documentclass[conference]{IEEEtran}
\IEEEoverridecommandlockouts

\usepackage{multirow}
\usepackage{tikz}
\usetikzlibrary{bayesnet}
\usepackage{siunitx}
\usepackage{subfig}
\usepackage{booktabs}
\usepackage{multirow}
\usepackage{siunitx} %
\usepackage{url}
\usepackage{bm}
\usepackage{tablefootnote} %

\usepackage{cite}
\usepackage{amsmath,amssymb,amsfonts}
\usepackage{algorithmic}
\usepackage{graphicx}
\usepackage{textcomp}
\usepackage{xcolor}
\def\BibTeX{{\rm B\kern-.05em{\sc i\kern-.025em b}\kern-.08em
    T\kern-.1667em\lower.7ex\hbox{E}\kern-.125emX}}

\definecolor{dynamics}{RGB}{217,95,2}
\definecolor{inference}{RGB}{27,158,119}
\definecolor{params}{RGB}{117,112,179}

\newcommand{\expectE}{\mathsf{E}}

\begin{document}
\onecolumn{

\large
\noindent Please cite this paper as: \newline
W. Ciezobka et al., ``FTMRate: Collision-Immune Distance-based Data Rate Selection for IEEE 802.11 Networks,'' 2023 IEEE 24th International Symposium on a World of Wireless, Mobile and Multimedia Networks (WoWMoM), Boston, MA, USA, 2023, pp. 242-251, doi: 10.1109/WoWMoM57956.2023.00039.

\normalsize
\vspace{1cm}
\begin{verbatim}

@INPROCEEDINGS{ciezobka2023ftmrate,
  author={Ciezobka, Wojciech and Wojnar, Maksymilian 
  and Kosek-Szott, Katarzyna and Szott, Szymon 
  and Rusek, Krzysztof},
  booktitle={2023 IEEE 24th International Symposium on a World 
  of Wireless, Mobile and Multimedia Networks (WoWMoM)}, 
  title={{FTMRate: Collision-Immune Distance-based Data Rate 
  Selection for IEEE 802.11 Networks}}, 
  year={2023},
  volume={},
  number={},
  pages={242--251},
  doi={10.1109/WoWMoM57956.2023.00039}}

\end{verbatim}

}
\twocolumn
\clearpage

\title{FTMRate: Collision-Immune Distance-based Data Rate Selection for IEEE 802.11 Networks %
\thanks{This research was funded  by the National Science Centre, Poland (DEC-2020/39/I/ST7/01457) and supported in part by PLGrid Infrastructure. For the purpose of Open Access, the author has applied a CC-BY public copyright licence to any Author Accepted Manuscript (AAM) version
arising from this submission.
\textit{(Corresponding author: Maksymilian Wojnar.)}}
}

\author{\IEEEauthorblockN{Wojciech Ciezobka, Maksymilian Wojnar, Katarzyna Kosek-Szott, Szymon Szott, Krzysztof Rusek}
\IEEEauthorblockA{
Institute of Telecommunications \\
AGH University of Krakow, Poland \\
Email: \{name.surname\}@agh.edu.pl
}
}

\maketitle

\begin{abstract}
Data rate selection algorithms for Wi-Fi devices are an important area of research because they directly impact performance. Most of the proposals are based on measuring the transmission success probability for a given data rate. In dense scenarios, however, this probing approach will fail because frame collisions are misinterpreted as erroneous data rate selection. We propose FTMRate which uses the fine timing measurement (FTM) feature, recently introduced in IEEE 802.11. FTM allows stations to measure their distance from the AP. We argue that knowledge of the distance from the receiver can be useful in determining which data rate to use. We apply statistical learning (a form of machine learning) to estimate the distance based on measurements, estimate channel quality from the distance, and select data rates based on channel quality. We evaluate three distinct estimation approaches: exponential smoothing, Kalman filter, and particle filter. We present a performance evaluation of the three variants of FTMRate and show, in several dense and mobile (though line-of-sight only) scenarios, that it can outperform two benchmarks and provide close to optimal results in IEEE 802.11ax networks.
\end{abstract}

\begin{IEEEkeywords}
Wi-Fi, 802.11, data rate selection, statistical learning
\end{IEEEkeywords}

\section{Introduction}

Rate selection algorithms for Wi-Fi devices are a productive area of research because they directly impact performance. 
The continued investigation in this area is caused by two factors.
First, new amendments to the IEEE 802.11 standard increase the selection space with more modulation and coding scheme (MCS) values, guard interval (GI) lengths, channel widths, etc.
For IEEE 802.11ax, Table~\ref{tab:80211axRates} presents the available data rates for just a single spatial stream. 
Since 802.11ax supports up to eight spatial streams, the number of data rates increases up to 1056. 
Second, the availability of new tools, such as those based on machine learning (ML) \cite{szott2022wifi}, can improve the rate selection performance.

We observe that many rate selection algorithms, even those applying ML, are based only on measuring the transmission success rate for a given MCS (Section~\ref{sec:soa}).
In dense scenarios, however, this probing approach will fail because collisions might be misinterpreted as erroneous rate selection.
Obviously, transmitters want to adapt their rates to the signal strength at the receiver (and not to the interference from other transmissions).
Therefore, an alternative closed-loop approach is to measure channel quality (such as the signal-to-noise ratio, SNR) at the receiver and send feedback to the transmitter.
In this vein, we propose FTMRate -- a closed-loop, context-aware rate selection algorithm which relies not on directly measuring channel quality, but on measuring distance using the fine timing measurement (FTM) feature introduced in IEEE 802.11-2016 \cite{80211-2016} 
(and recently extended in 802.11az). 
With FTM, stations can estimate their distance from the AP using round-trip time (RTT) measurements. 
Knowing the distance from the receiver can be useful in determining which rate to use. 
However, this brings about the following challenges:
\begin{itemize}
    \item estimating the distance from FTM measurements,
    \item estimating channel quality from the distance,
    \item selecting MCS values based on the channel quality.
\end{itemize}
Fortunately, distance estimation need not be as precise as for positioning (the main use case of FTM) since rate selection is more robust to measurement errors.
We claim that estimating distance allows estimating the channel quality described as the expected SNR at the receiver side, 
provided that we know the channel model. 
Still, there remains the problem of considering slow (shadowing) and fast (multi-path) channel fading.
For the former, in this study, we only target line-of-sight (LOS) scenarios (to evaluate the validity of the approach) and leave non-LOS as future work.
Regarding the latter, we ignore adapting to multi-path fading, which in the timescale of a frame transmission is impractical.
Finally, since the dependency between SNR and successful transmission probability is known for each MCS \cite{krotov2020}, it is possible to select the appropriate rate for each transmission.

The overall goal of this work is to evaluate whether using FTM can improve rate selection.
After an overview of the state of the art (Section~\ref{sec:soa}), we provide the following contributions:
\begin{itemize}
    \item we succinctly describe the operation of FTM and discuss overhead issues in Section~\ref{sec:rttDistance},
    \item we describe our proposal, FTMRate, and its model in Section~\ref{sec:ourProposal}, 
    \item we present a performance evaluation of three variants of FTMRate and three benchmarks in several dense and mobile scenarios (Section~\ref{sec:Scenarios}).
\end{itemize}
Finally, we conclude the paper %
in Section~\ref{sec:conclusions}.

\begin{table*}[t!]
\sisetup{detect-weight,
         mode=text    %
         }
\caption{IEEE 802.11ax transmission rates for a single spatial stream}
\begin{tabular}{lll|lll|lll|lll|lll}
\hline
\multicolumn{1}{c}{\multirow{3}{*}{\textbf{\begin{tabular}[c]{@{}c@{}}MCS\\ index\end{tabular}}}} & \multicolumn{1}{c}{\multirow{3}{*}{\textbf{\begin{tabular}[c]{@{}c@{}}Modulation \\ type \end{tabular}}}} & \multicolumn{1}{c}{\multirow{3}{*}{\textbf{\begin{tabular}[c]{@{}c@{}}Coding\\ rate\end{tabular}}}} & 
\multicolumn{11}{c}{\textbf{Data rate (Mbit/s) for different channel widths [MHz] and GIs [\SI{}{\micro\second}]}} \\
\multicolumn{1}{c}{}   & \multicolumn{1}{c}{}   & \multicolumn{1}{c}{} & \multicolumn{3}{c}{\textbf{20 MHz}} & \multicolumn{3}{c}{\textbf{40 MHz}}   & \multicolumn{3}{c}{\textbf{80 MHz}}   & \multicolumn{3}{c}{\textbf{160 MHz}}  \\
\multicolumn{1}{c}{}   & \multicolumn{1}{c}{}   & \multicolumn{1}{c}{} & 
\multicolumn{1}{|c}{\textbf{\SI{3.2}{\micro\second}}} & \multicolumn{1}{c}{\textbf{\SI{1.6}{\micro\second}}} & \multicolumn{1}{c}{\textbf{\SI{0.8}{\micro\second} }} & 
\multicolumn{1}{|c}{\textbf{\SI{3.2}{\micro\second}}} &\multicolumn{1}{c}{\textbf{\SI{1.6}{\micro\second}}} & \multicolumn{1}{c}{\textbf{\SI{0.8}{\micro\second}}} & 
\multicolumn{1}{|c}{\textbf{\SI{3.2}{\micro\second}}} &\multicolumn{1}{c}{\textbf{\SI{1.6}{\micro\second}}} & \multicolumn{1}{c}{\textbf{\SI{0.8}{\micro\second}}} & 
\multicolumn{1}{|c}{\textbf{\SI{3.2}{\micro\second}}} &\multicolumn{1}{c}{\textbf{\SI{1.6}{\micro\second}}} & \multicolumn{1}{c}{\textbf{\SI{0.8}{\micro\second}}} \\
\hline
0 & BPSK 	 	& 1/2 		& 7.3  & 8.1  & 8.6  		& 14.6 & 16.3  & 17.2  		& 30.6 & 34.0  & 36.0  		& 61.3 & 68.1   & 72.1   \\
1 & QPSK 	 	& 1/2 		& 14.6 & 16.3 & 17.2 		& 29.3 & 32.5  & 34.4  		& 61.3 & 68.1  & 72.1  		& 122.5 & 136.1  & 144.1  \\
2 & QPSK 	 	& 3/4 		& 21.9 & 24.4 & 25.8 		& 43.9 & 48.8  & 51.6  		& 91.9 & 102.1 & 108.1 		& 183.8 & 204.2  & 216.2  \\
3 & 16-QAM 	 	& 1/2 		& 29.3 & 32.5 & 34.4 		& 58.5 & 65.0  & 68.8  		& 122.5 & 136.1 & 144.1 	& 245.0 & 272.2  & 288.2  \\
4 & 16-QAM 	 	& 3/4 		& 43.9 & 48.8 & 51.6 		& 87.8 & 97.5  & 103.2 		& 183.8 & 204.2 & 216.2 	& 367.5 & 408.3  & 432.4  \\
5 & 64-QAM 	 	& 2/3 		& 58.5 & 65.0 & 68.8 		& 117.0 & 130.0 & 137.6 	& 245.0 & 272.2 & 288.2 	& 490.0 & 544.4  & 576.5  \\
6 & 64-QAM 	 	& 3/4 		& 65.8 & 73.1 & 77.4 		& 131.6 & 146.3 & 154.9 	& 275.6 & 306.3 & 324.4 	& 551.3 & 612.5  & 648.5  \\
7 & 64-QAM 		& 5/6 		& 73.1 & 81.3 & 86.0 		& 146.3 & 162.5 & 172.1 	& 306.3 & 340.3 & 360.3 	& 612.5 & 680.6  & 720.6  \\
8 & 256-QAM 	& 3/4 		& 87.8 & 97.5 & 103.2 		& 175.5 & 195.0 & 206.5 	& 367.5 & 408.3 & 432.4 	& 735.0 & 816.7  & 864.7  \\
9 & 256-QAM 	& 5/6 		& 97.5 & 108.3 & 114.7 		& 195.0 & 216.7 & 229.4 	& 408.3 & 453.7 & 480.4 	& 816.6 & 907.4  & 960.7  \\
10 & 1024-QAM	& 3/4 		& 109.7 & 121.9 & 129.0 	& 219.4 & 243.8 & 258.1 	& 459.4 & 510.4 & 540.4 	& 918.8 & 1020.8 & 1080.9 \\
11 & 1024-QAM	& 5/6 		& 121.9	& 135.4 & 143.4 	& 243.8 & 270.8 & 286.8 	& 510.4 & 567.1 & 600.5 	& 1020.8 & 1134.2 & 1201.0 \\
\hline
\end{tabular}
\label{tab:80211axRates}
\end{table*}

\section{State of the art}
In this section, we review both non-ML and ML-based rate selection methods and state the novelty of our approach.
\label{sec:soa}
\subsection{Non-ML-based Methods}

Numerous studies have been carried out on rate adaptation methods \cite{yin2020mac}.
Minstrel \cite{fietkau2020minstrel} is a well-known sampling-based method:
it calculates the transmission success ratio (based on received acknowledgment frames, ACKs) and %
performs %
moving average calculations to process the success history of each rate. Then, the rates with the best throughput, second-best throughput, highest success probability, and lowest base rate are considered for transmission. %
Due to its good performance, 
Minstrel is the default rate adaptation algorithm in the Linux kernel and we use it as a baseline in Section~\ref{sec:Scenarios}.

\subsection{ML-based Methods}
ML-based rate control is mostly supervised or reinforcement learning-based, often using link quality indicators %
as inputs.

\subsubsection{Supervised Learning}
In \cite{punal2013rfra}, random forests (RFs) are applied in IEEE 802.11p networks to select transmission rates based on the predicted probability of successful transmission. SNR samples are used to characterize the propagation environment, and position/speed information (obtained from GPS) is used to increase the prediction accuracy. 
In \cite{wang2013dynamic}, a multilayer perceptron artificial neural network is adopted to model the correlation function between the optimal number of consecutive failed or successful uplink transmissions and the set of different traffic metrics (number of contending nodes, channel conditions measured as the bit error rate (BER), and traffic intensity). 
Furthermore, \cite{li2020practical} implements an artificial neural network that learns the mutual influence of rates, throughput, and channel quality. The data rate is selected based on the observed congestion level (measured with the A-MPDU service time). 
Furthermore, \cite{kurniawan2018machine} uses RFs to classify the channel type (residential or office) to select the appropriate MCS while \cite{hussien2021towards} treats rate selection as a  multi-label multi-class classification problem and uses deep neural networks.

\subsubsection{Reinforcement Learning}
Several works implement \textit{stochastic approaches}.
In \cite{joshi2008sara}, stochastic learning automata-based rate adaptation is proposed to guarantee a given packet success rate by observing successful transmission attempts. 
In \cite{karmakar2016dynamic}, a stochastic multi-armed bandit-based distributed algorithm is proposed that explores different configuration options (e.g., channel bandwidth, MCS) and observes their impact on network performance under varying channel conditions. The frame success ratio (FSR) is used as a reward and the optimal configuration (i.e., with the highest FSR) is used to calculate the regret. 
In \cite{gupta2018low}, Thompson sampling (TS)~\cite{thompson1933likelihood} is used to model the acknowledgment probability for each MCS and maximize throughput. 
In \cite{krotov2020}, the transmission rate is adjusted based on channel quality, represented by the signal-to-interference ratio (SINR) and the transmission power. Additionally, a particle filter is used to estimate the channel quality (SINR).  
Latent TS is proposed in \cite{saxena2021reinforcement}, where SINR is modeled as a probability distribution %
and updated based on the received feedback (ACK/NACK). Additionally, the SINR probability distribution is relaxed at each time step with a smoothing function to adapt to fading channels. 
Graphical optimal rate sampling is proposed in \cite{combes2018optimal}. The authors assume the throughput to be a unimodal function of the selected (rate, MIMO mode) pair. They show that graphical unimodality can be used to efficiently learn and track the best transmission rates.
Other recent approaches include using batched Thompson sampling \cite{chen2022adaptive} and improving Minstrel performance with shallow neural networks \cite{onasis2022improving}.

In terms of \textit{more complex ML approaches}, the authors of \cite{yano2022study} apply Q-learning (QL) for rate adaptation in overlapping networks. 
In \cite{cho2021reinforcement}, packet timeouts are used to train the agent. Choosing an MCS is an action, and the resulting contention window (CW) size is a state. 
Deep Q-learning (DQL) is used in
\cite{chen2021experience} to adapt to the current link quality and channel conditions. 
Rate selection features (MCS, MIMO mode, channel width) are treated as different coordinates of a 3D maze. Each 3D cell in the proposed maze represents a rate. Then, a DRL model is used, in which the action space ``includes two moving directions along three dimensions'', the state includes six elements (channel width, MIMO mode, MCS, sub-frame loss, RSSI, service time ratio), and the resulting goodput is treated as the reward.

\subsection{Novelty of Approach}
Our proposal, FTMRate, differs from other state-of-the-art approaches based on machine learning, e.g., TS and deep reinforcement learning (DRL)~\cite{mnih2013playing}. These general methods assume no knowledge about how the channel behaves and the only feedback comes from a frame transmission, either successful or not.
While an FTM measurement could be yet another signal from the environment (called the `context' in RL literature) that would improve the accuracy of these methods, it would also pose some issues. In the case of TS, its simplicity is lost when contextual bandits are introduced since we have no simple posterior formula. 
Meanwhile, a general DRL agent could exploit FTM information, but such an approach is computationally expensive, and the DRL agent needs to learn what is already known, i.e., have prior knowledge of how the communication channel works. In this paper, working in the paradigm of context-aware communication, we fully embrace knowledge of the wireless channel to build a simple yet efficient agent based on probabilistic modeling.

\section{Distance Estimation Using Fine Timing Measurement}

\label{sec:rttDistance}

\begin{figure}[t!]
\centering
\includegraphics[width=0.45\textwidth]{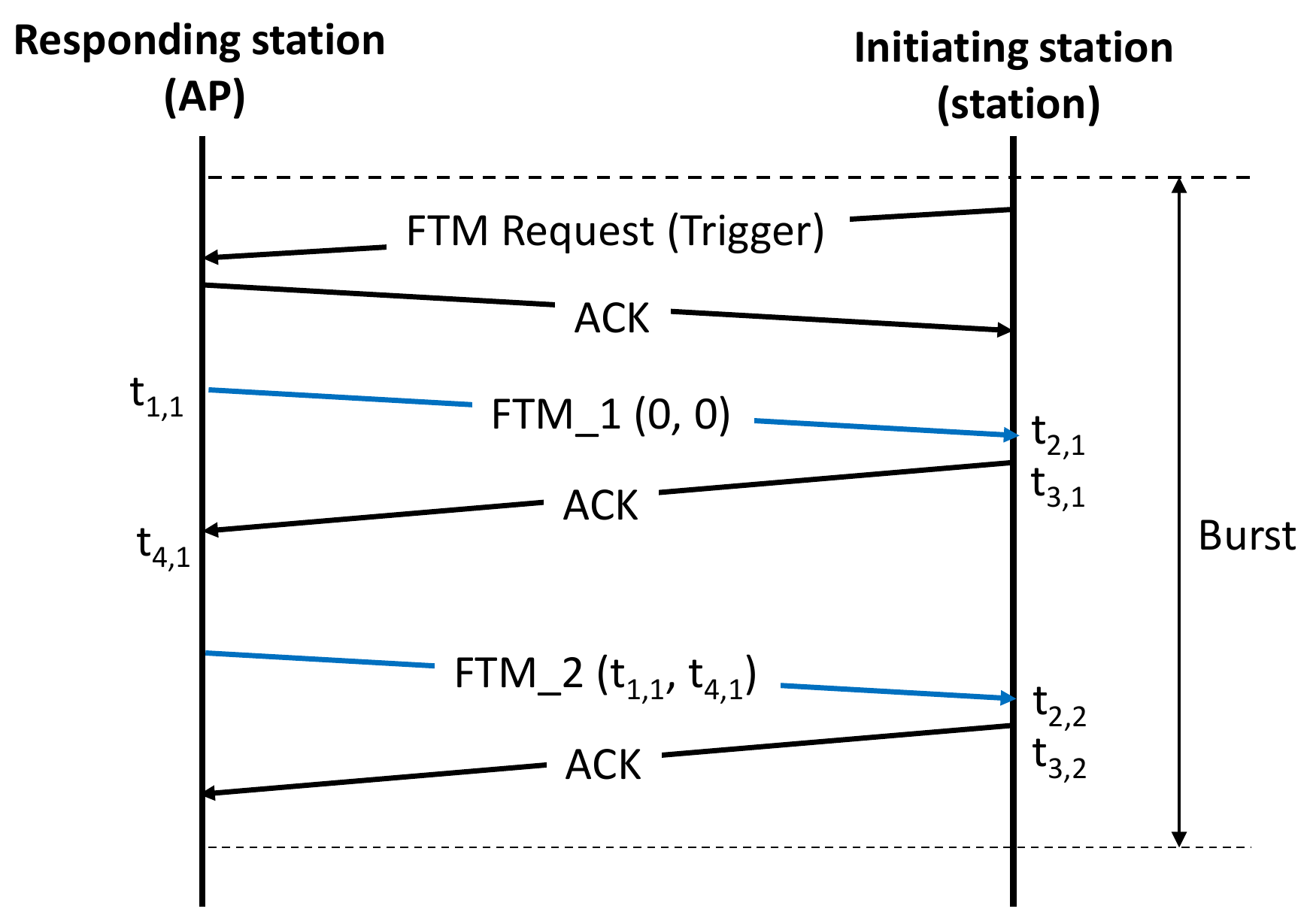}
\caption{FTM operation for the shortest possible burst \cite{ieee802112020}.}
\label{fig:ftm}
\end{figure}

FTM operation is based on a burst of frame exchanges, as shown in Fig.~\ref{fig:ftm}. First, the initiating station sends a trigger (FTM Request) to initiate the measurement procedure. Then, FTM frames and ACKs are exchanged between the responding station and the initiating station. The goal is for the initiating station to obtain four timestamps for the $i$-th frame exchange ($t_{1,i}$ to $t_{4,i}$) so as to compute the RTT:
\begin{equation}
    RTT_i = (t_{4,i} - t_{1,i}) - (t_{3,i} - t_{2,i}),
\end{equation}
where $t_{1,i}$ is the time at which the AP transmitted the $i$-th FTM frame, $t_{4,i}$ is the time at which the acknowledgment of the $i$-th FTM frame was received by the AP, $t_{2, i}$ is the time at which the $i$-th FTM frame was received by the initiating station, and $t_{3, i}$ is the time at which the station transmitted the acknowledgment of the correct reception of the $i$-th FTM frame to the AP \cite{ieee802112020}. 
FTM requires $n$ frame exchanges to calculate $n-1$ RTTs.
Therefore, the minimum number is $n=2$, as presented in Fig.~\ref{fig:ftm}, where a single RTT is measured in the burst, and the $t_{1,1}$ and $t_{4,1}$ timestamps are transferred by the AP to the initiating station in the second FTM frame (\texttt{FTM\_2}).
The $i$-th distance estimation ($\rho^{RTT}_i$) can be calculated from the $i$-th RTT as
\begin{equation}
    \rho^{RTT}_i = \frac{RTT_i}{2} c, 
\end{equation}
where $c$ is the speed of light.

The signalling overhead required to perform an FTM operation depends, i.a., on the size of the burst. 
Zubow et al. have found that ``in scenarios
with strong multipath environments, e.g., indoors, there is no gain from using
larger channel bandwidth and higher number of measurements''  \cite{zubow2022ftm}.
Therefore, the airtime consumed by the shortest possible FTM burst (Fig.~\ref{fig:ftm}) consists of transmitting an FTM request frame, two FTM frames, and three ACK frames.
Table~\ref{tab:ftmduration} presents the duration of the shortest FTM burst. 
FTM frame lengths were taken from the implementation provided in \cite{zubow2022ftm} and the frame durations -- from the WiFi AirTime Calculator\footnote{\url{https://gjermundraaen.com/thewifiairtimecalculator/}}.
The table presents two cases: the main case when FTM frames are transmitted as 802.11a frames at 6 Mb/s and an alternative case when these frames are transmitted at the fastest rate available (802.11ax's 143.4 Mb/s).
The latter is hypothetical and is provided to show the lower bound for FTM airtime duration.

Clearly, we see that reducing FTM-induced overhead cannot be achieved by using higher rates: the burst airtimes are similar on account of the frame header transmission durations, which are always sent at the lowest control rate.
However, a promising solution to solve the overhead problem would be to perform FTM out-of-band. 
IEEE 802.11be provides multi-link operation (MLO) and it is easy to envision a case when the main communication between stations and APs is done in the 5 or 6~GHz bands, while FTM bursts are relegated to the 2.4~GHz band.
In the following, we focus on the rate selection performance improvements achieved by estimating the distance with FTM and leave overhead considerations as future work.

\begin{table}[t!]
\caption{Airtime duration of the shortest possible FTM burst (Fig.~\ref{fig:ftm}). The total burst airtime consists of the FTM frame transmission durations, three ACK transmissions, and three SIFS (16~\si{\micro\second}) periods. In  802.11a, the FTM frames are transmitted at 6 Mb/s, in 802.11ax -- at 143.4 Mb/s.
We assume transmission on a 20~MHz channel with a single spatial stream and a guard interval of 0.8~\si{\micro\second}.}
\label{tab:ftmduration}
\centering
\begin{tabular}{@{}llll@{}}
\toprule
\multicolumn{1}{c}{\multirow{2}{*}{Frame}} & \multicolumn{1}{c}{\multirow{2}{*}{Length {[}B{]}}} & \multicolumn{2}{c}{Duration {[}\si{\micro\second}{]}}    \\  
\multicolumn{1}{c}{}                       & \multicolumn{1}{c}{}                                & 802.11a & 802.11ax \\ \cmidrule(r){1-4}
FTM Request                                & 42                                                  & 80               & 58                    \\
FTM\_1                                     & 66                                                  & 112              & 58                    \\
FTM\_2                                     & 48                                                  & 88               & 58                    \\
ACK                                        & 14                                                  & 44               & 32                    \\
\multicolumn{2}{r}{Total burst airtime:}                                                                 & 460              & 318                   \\ \bottomrule
\end{tabular}
\end{table}

\section{FTMRate Proposal}
\label{sec:ourProposal}

We begin our description of FTMRate by listing several assumptions, which we consider throughout this study:
\begin{itemize}
\item
  only line-of-sight communication (an indoor scenario),
\item
  static or low-mobility (nomadic) stations,
\item
  only MCS-based rate selection,
\item
  an exponential Gaussian error model of FTM measurements based on \cite{zubow2022ftm},
\item
  a log-distance path loss model (with known parameters) and Nakagami multipath fading,
\item
  perfect SNR to MCS mapping,
\item
  fixed transmission power level,
\item
  an FTM measurement rate of once every 0.5 s (2 Hz), performed out-of-band.
\end{itemize}
These assumptions are stringent, but they allow an initial study of FTMRate performance.
We leave addressing these assumptions as future work.
Nevertheless, the chosen channel and network characteristics (indoor, low mobility, high density) are comparable to a fully occupied large concert hall, where network connectivity becomes an issue.

In FTMRate, each 802.11 station runs an independent ML agent that selects rates (MCS values).
In the construction of this agent, we use probabilistic modeling, which is crucial as FTM measurements are inaccurate and 802.11 is far from being a simple system.
Probabilistic modeling means that every unknown is modeled as a random variable with some probability distribution.
Certain complicated aspects of the system are also abstracted by statistical learning models.
Under some mild assumptions, we can reduce the measurement error by including past measurements in the model.
The net result of such an approach is the distribution of the possible rate for each MCS at every point in time.
This distribution allows the selection of the optimal MCS and to control the FTM probing rate to reduce overhead.

We model FTMRate as follows.
Each station measures its distance from the AP at time $t$ using FTM to measure the RTTs.
The true distance is denoted as $\rho_t$ and its noisy reading as $\rho_t^{RTT}$.
The channel model is assumed to be known and has the form of the conditional distribution  %
of SNR given the distance between the sender and the receiver\footnote{Channel model estimation and prediction is left for future study.}: $p_{\bm \theta_c}(\gamma|\rho)$, where $\gamma$ is the SNR and $\bm \theta_c$ is a vector of the channel model parameters.
In the following, the subscript denotes parameters of a function, e.g., $p_{\bm \theta_c}(\cdot):=p(\bm \theta_c, \cdot)$.
A frame transmission is successful with probability $\xi=s_{\bm \theta_s}(\gamma,\mu)$,
where 
$\mu$ is the MCS used for the transmission and $s_{\bm \theta_s}$ is the CDF of the sinh-arcsinh distribution.
Since distance measurements are noisy, a smoothing filter is applied to obtain a more accurate estimate of the actual distance.

\begin{figure}[t!]

  \begin{center}
\begin{tikzpicture}

\tikzset{latent/.append style={draw=dynamics,text=dynamics}}
	\node[latent]                         (nutm1)      {{$\nu_{t-1}$}};
	\node[latent, right=1cm of nutm1]                         (nut)      {$\nu_{t}$};
	\node[latent, below=1cm of nutm1]    (rhotm1)      {$\rho_{t-1}$};
	\node[latent, right=1cm of rhotm1]    (rhot)      {$\rho_{t}$};
	\node[obs, right=1cm of rhot]    (obs)      {$\rho^{RTT}_{t}$};
 
 \tikzset{latent/.append style={draw=inference,text=inference}}
 \tikzset{const/.append style={text=params}}
 
	\node[const, above right=1cm of nut]    (parsn)      {$\sigma_\nu$};
	\node[const, right=1cm of parsn]    (parsr)      {$\sigma_\rho$};
	\node[const, right=1cm of parsr]    (tau)      {$\tau$};
	\node[latent, below=1cm of rhot]    (snr)      {$\gamma$};
	\node[const, above right=0.5cm of snr]    (tc1)      {$\bm\theta_c$};%
    
    \draw[dashed, thin] (-2,-2.5) -- (6.5,-2.5);
    \node[text width=1.5cm,right=0.5cm of obs]    (mob)      {\textcolor{dynamics}{mobility dynamics}};	
    \node[text width=1.5cm, below=0.5cm of mob]    (inf)      {\textcolor{inference}{rate inference}};	

	\node[latent, below=1cm of snr]    (s)      {$\xi$};

	\node[latent, below=1cm of s]    (rate)      {$\lambda$};
	
	\node[const,left=1cm of snr] (mcs) {$\mu$};
	\node[const,below left=1.4142cm of mcs] (lambdamax) {$\lambda^{\mathrm{tx}}_{\mu}$};
	\node[const, below=1cm of mcs]    (ts1)      {$\bm\theta_s$};

 \edge {nutm1}{nut} ; %
 \edge {rhotm1}{rhot} ; %
 \edge {nutm1}{rhot} ; %
 \edge {rhot}{obs} ; %
  \edge {parsn}{nut} ; %
  \edge {tau}{nut} ; %
  \edge {tau}{rhot} ; %
    \edge {parsn}{rhot} ; %
  \edge {parsr}{rhot} ; %
  \edge {rhot}{snr} ; %
  \edge {snr}{s};
  \edge {tc1}{snr};
  \edge {s}{rate};
  \edge {mcs}{ts1};
  \edge {ts1}{s}
  \edge {mcs}{lambdamax}
  \edge {lambdamax}{rate}

\end{tikzpicture}
\end{center}
  \caption{Graphical representation of the proposed model that relates distance observations with a hidden state and the rate achievable at a given MCS. Each arrow represents statistical dependence between random variables (circled) and parameters (\textcolor{params}{uncircled}). The graph comprises two parts: \textcolor{dynamics}{station mobility dynamics} and \textcolor{inference}{rate inference}.\label{fig:graphical}}
\end{figure}
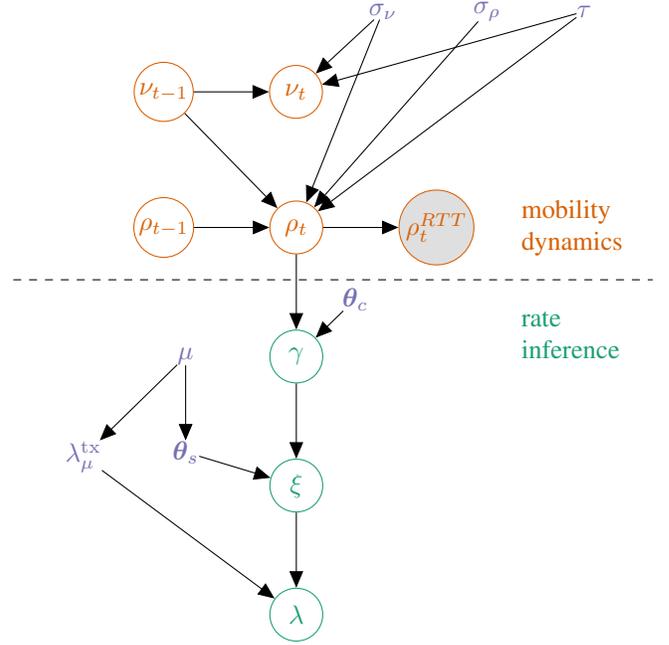

The graphical model we use is depicted in Fig.~\ref{fig:graphical}:
the top part represents mobility dynamics (used to smooth the noisy FTM readings), the bottom part represents MCS inference.
In the following, we describe both parts in detail.
The notation is summarized in Table~\ref{tab:notation}.

\begin{table}[t!]
  \centering
  \caption{Notation used.}
    \begin{tabular}{cp{20em}}
    \toprule
    Name & Definition \\
    \hline
    $\alpha$& level exponential smoothing weight\\
    $\beta$& trend exponential smoothing weight\\
    $\gamma$& SNR \\
    $\delta_{t}$&  trend transition noise\\
    $\epsilon_t$&  distance measurement error \\
    $\varepsilon_t$&  level transition noise \\
    $\bm\theta_c$&  channel model parameters\\    
    $\bm\theta_s $& parameters of \emph{sinh-arcsinh} normal distribution \\
    $p_{\bm\theta_c}(\gamma|\rho)$& channel model \\
    $\lambda$& expected rate \\
    $\lambda^{\mathrm{tx}}_{\mu}$& transmission rate for the $\mu$-th MCS \\
    $\mu$& current MCS \\
    $\nu$& radial velocity \\  
    $\nu_{t}$& trend (velocity like) at discrete time $t$\\
    $\xi$& probability of successful transmission \\
    $\rho$& true distance \\
    $\rho^{RTT}_{t}$ & noisy reading of true distance from the AP \\
    $\rho_{t}$& distance from the AP at discrete time $t$\\
    $\sigma_\nu$& variation of velocity \\
    $\sigma_\rho$ & variation of distance \\
    $\tau$& time between measurements or the time since the latest measurement \\    
    $s_{\bm \theta_s}$ & CDF of the sinh-arcsinh distribution \\
    \bottomrule
    \end{tabular}%
  \label{tab:notation}%
\end{table}%

\subsection{Mobility Dynamics}

Upon each measurement at discrete time $t$, the agent obtains a noisy reading $\rho^{RTT}_t$ of the true distance between the transmitter and the receiver.
We may reduce the distance uncertainty by the sequential filtering of observations. 
In particular, we assume that the true distance $\rho$ evolves with the local lineal trend $\nu$ and both obey the following stochastic differential equations~\cite{karatzas1991brownian}:
\begin{align}\label{eq:sde}
\begin{cases}
     \mathrm{d}\nu&=\sigma_\nu\mathrm{d}W_1\\
    \mathrm{d}\rho&=\nu \mathrm{d}t+\sigma_\rho\mathrm{d}W_2,
\end{cases}    
\end{align}
where $\mathrm{d}W_1$ and $\mathrm{d}W_2$ are independent Wiener processes and the observations are assumed to be the true distance with noise~$\epsilon_t$: $\rho^{RTT}_t=\rho_t+\epsilon_t$, whose distribution could be either centered normal with known variance or exponentially modified Gaussian, also with known parameters~\cite{zubow2022ftm}. Furthermore, $\sigma_\nu\in \mathbb{R_+}$ is the velocity variation and $\sigma_\rho \in \mathbb{R_+}$ is the distance variation.

This approach is inspired by the physics of Brownian motion.
We assume that each station has a latent radial velocity $\nu$ that is subject to random changes.
The variation of these changes is controlled by $\sigma_{\nu}$ and setting $\sigma_{\nu}=0$ gives us the dynamics of~\cite{krotov2020}.
As in physics, the infinitesimal change of distance $\rho$ depends on the current value of $\nu$, however, since this is just an approximation, we capture other effects, which change $\rho$, as additive noise.
The amount of this noise is controlled by the $\sigma_\rho$ parameter.

The particular choice for random processes representing noise is also an approximation, as it allows for negative distances.
In practice, this is not a problem, and the analytical solution is a great advantage of such a model.

The solution to~\eqref{eq:sde} is known as the Ornstein--Uhlenbeck process, a continuous-time stochastic process.
In practice,  the process is observed (sampled)
at discrete times $t=t_{2,i+1}$, which yields the following discrete linear dynamical system:
\begin{align}
\begin{cases}
     \nu_{t+1}&=\nu_{t}+\delta_{t},\\
    \rho_{t+1}&=\rho_{t}+\nu_{t}\tau+ \varepsilon_t,\\
    \rho^{RTT}_t&=\rho_t+\epsilon_t,
\end{cases}
\end{align}
where $\tau$ is either the time between measurements $\tau=(t_{2,i+2}-t_{2,i+1})$ or the time since the last measurement, and both process (transition) noises, $\delta_{t}$ and $\varepsilon_t$, have a joint multivariate normal distribution:
\begin{equation}
    \begin{bmatrix}
        \delta_{t}\\ \varepsilon_t
    \end{bmatrix}\sim \mathcal{N}\left(%
    \begin{bmatrix}
    	0\\0%
    \end{bmatrix},
    \left[\begin{matrix}\sigma_{\nu}^{2} \tau & \frac{\sigma_{\nu}^{2} \tau^{2}}{2}\\\frac{\sigma_{\nu}^{2} \tau^{2}}{2} & \tau \left(\frac{\sigma_{\nu}^{2} \tau^{2}}{3} + \sigma_{\rho}^{2}\right)\end{matrix}\right]
    \right).
\end{equation}

\begin{figure*}[]
    \centering
    \subfloat[Measuring and smoothing distance (left) and mapping to SNR (right) using~\eqref{eq:dist_to_snr}.]{\includegraphics[width=1.0\linewidth]{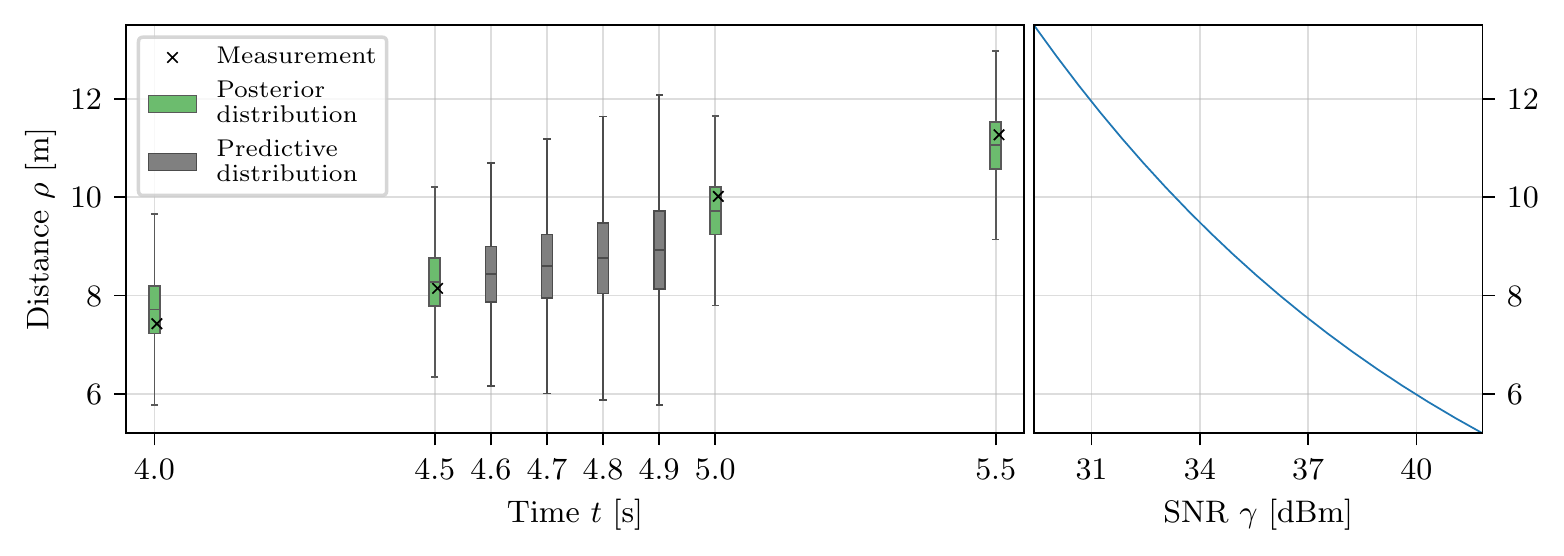}}
    
    \medskip
    
    \raggedright
    \subfloat[Estimating SNR (left) and mapping to MCS (right), where the dashed lines indicate the maximum rate at a given MCS.]{\includegraphics[width=1.0\linewidth]{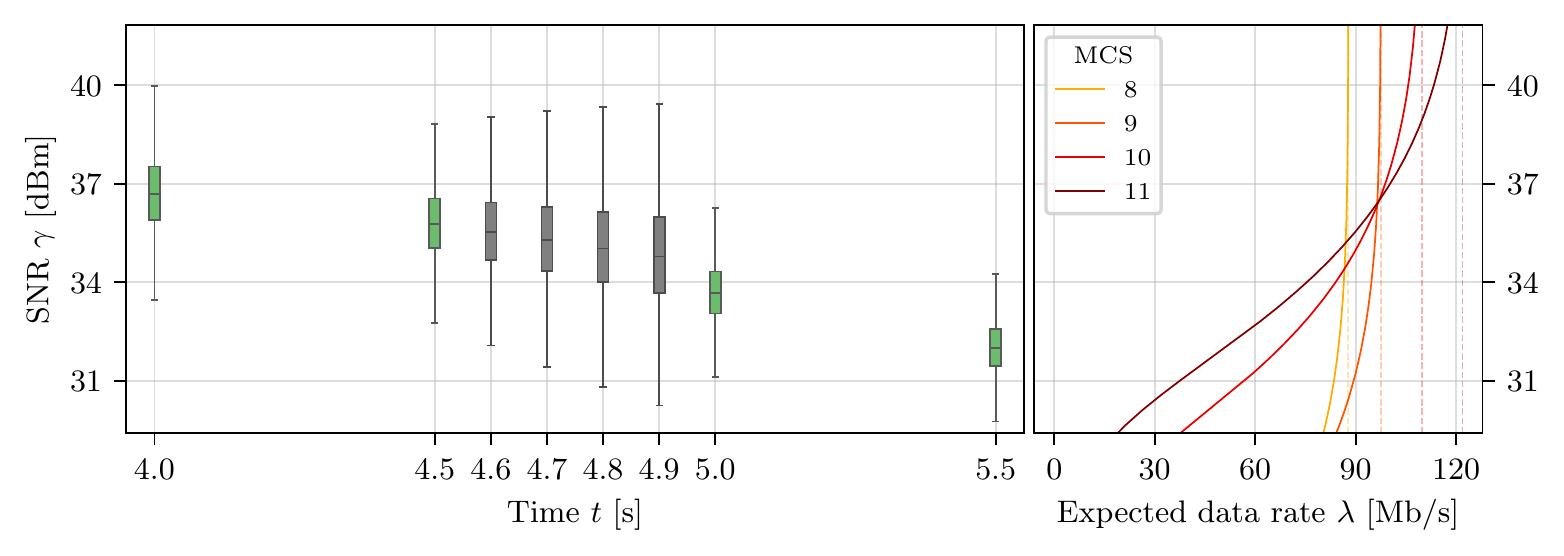}}
    
    \medskip
    
    \subfloat[Selecting MCS based on expected data rate, represented as boxplots for each time interval.]{\includegraphics[width=0.7067\linewidth]{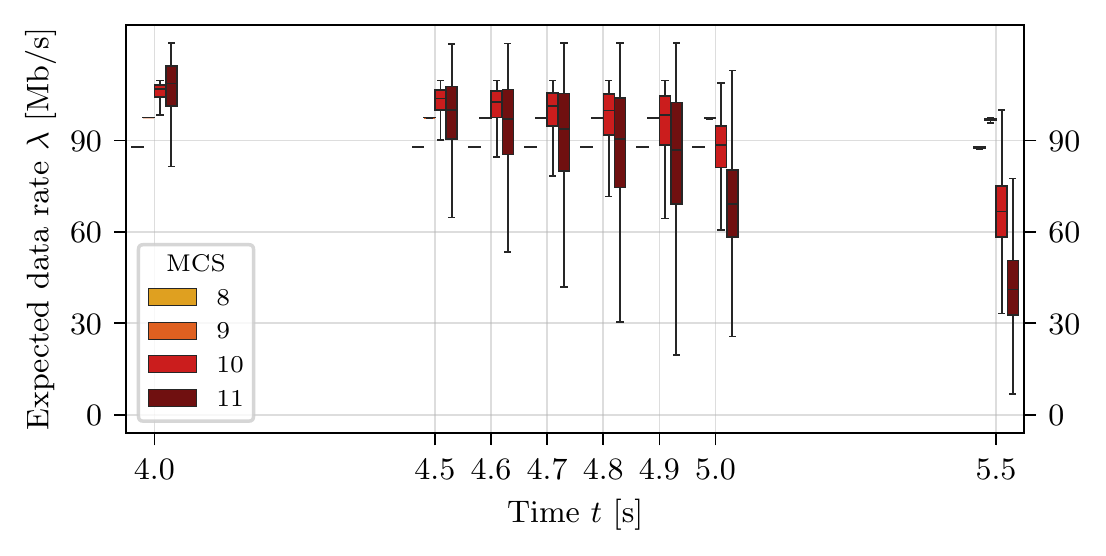}
    \label{fig_collage_selecting_mcs}}
    
    \caption{The FTMRate method broken into subalgorithms in a simple example (using the Kalman filter approach). 
    This example is for a station transmitting frames while moving away from the AP with a velocity of \SI{2}{\meter/\second}. Figures on the left show how distributions evolve over time, while the ones on the right present corresponding bijective transformations between quantities on the (vertically adjacent) time evolution figures. 
    }
    \label{fig_collage}
\end{figure*}

If FTM were costless and instantaneous, an agent could trigger it upon each frame transmission.
However, since it introduces overhead, we must keep the measurement rate as low as possible to protect link capacity.
The key to achieving high throughput is a discretized continuous time model that can be used at a random time while being updated at discrete times.
The internal state $(\rho,\nu)$ can be inferred from observations at discrete time points, while a continuous model extrapolates dynamics of frame transmission times that occur between measurements, as shown in Fig.~\ref{fig_collage}. 
We observe that the uncertainty increases over time, and a new measurement makes the estimate more certain.
The trend component $\nu$ increases the inertia of the model, thus allowing longer times between measurements.
Inference can be done using any of the popular methods, such as Kalman filter (KF) or particle filter (PF).
The selected method is dictated by the assumption about the distribution of $\epsilon$. 
For a normal distribution, we have a linear Gaussian state space model and the Kalman filter is an analytical solution for posterior inference~\cite{murphy2012machine}.
Any other distribution requires a more general or approximate method, and in such a case we use a particle filter~\cite{chopin2020introduction}.
Note that it is important to correctly estimate distance uncertainty, as the subsequent transformations are highly non-linear, as shown in Fig.~\ref{fig_collage}.
A Taylor series expansion (not shown here) confirms that the variance of distance transforms into a bias in the expected rate.
Thus, by measuring and reducing distance uncertainty, we get more accurate estimates of the expected rates.

Having said that, we also use double exponential smoothing (ES) with linear trend~\cite{hyndman2008forecasting} as the third method of inference, serving as a baseline. 
ES, 
being a Holt linear model \cite{hyndman2008forecasting}, estimates $\rho$ as $\rho_{t}=l_{t-1}+\tau s_{t-1}$, where the state (decomposed into trend $s$ and level $l$) is estimated with simple filtration~\cite{hyndman2008forecasting}:
\begin{align}
\begin{cases}
    l_{t+1}=&\alpha\rho_t^{RTT}+(1-\alpha)(l_t+v_t),\\
    s_{t+1}=&\beta(l_{t+1}-l_{t})+(1-\beta)s_t,
\end{cases}
\end{align}
where $\alpha,\beta\in (0,1)$ are the model parameters. 
This approach is simple and fast; however, it requires a constant sampling rate because it lacks a proper probabilistic interpretation for continuous time.

In summary, we study FTMRate in three inference variants: Kalman filter, particle filter, and exponential smoothing.
Such inference is required because during the time interval between consecutive FTM measurements we need to deduce the distance from filtered past discrete measurements. 
We then parameterize the predictive distribution of these measurements using one of our filters (PF, KF, ES).
The choice of method is influenced by the available computational resources. The weights of filtration with a Kalman filter or exponential smoothing are both comparable with the computational weight of rate adaptation using TS. However, the cost of performing filtration with a particle filter is much higher, depending strictly on the number of particles. 
Nonetheless, all our methods are conceptually simple and should be possible to implement even on simple devices with low computing power.

\subsection{Rate Inference}
Formally, the problem we are dealing with is a Markov decision process because the dynamics of~\eqref{eq:sde} form a Markov process.
However, since the state transition is independent of the action taken by the agent (i.e., the chosen MCS), in our proposal, the MCS is selected according to
\begin{equation}\label{eq:opt}
    \mu^*(\rho)=\arg\max_\mu \expectE_{\gamma\sim p_{\bm\theta_c}(\gamma|\rho)}\lambda^{\mathrm{tx}}_{\mu} s_{\bm \theta_s}(\gamma,\mu),
\end{equation}
where $\lambda^{\mathrm{tx}}_{\mu}$ is the transmission rate for the $\mu$-th MCS.
This selection process includes multiple steps detailed below and exemplified in Fig.~\ref{fig_collage}. 

First, we measure and smooth the distance. Then, we map the distance $\rho_t$ to SNR $\gamma_t$ using the following log-distance channel model:
\begin{equation}\label{eq:dist_to_snr}
    \gamma(\rho)=\gamma_0 - \left(L_0 + 10E\log_{10}(\rho)\right),
\end{equation}
where $E$ is the path loss exponent, $\gamma_0$ -- the reference SNR, and $L_0$ -- a reference path loss measured at a distance of $\SI{1}{\meter}$.
Therefore, the parameters of the channel model used are $\bm\theta_c=(\gamma_0,L_0,E)\in \mathbb{R}^2 \times \mathbb{R_+}$ and the resulting distributions are shown in Fig.~\ref{fig_collage}b.
Next, we compute the probability of successful transmission $\xi$ modeled with a CDF of the \emph{sinh-arcsinh} normal distribution~\cite{sinh-arcsinh} whose four parameters $\bm\theta_s\in\mathbb{R}^2\times\mathbb{R_+}^2$ are estimated from offline simulations.
Finally, we scale the rate at each MCS ($\lambda^{\mathrm{tx}}_{\mu}$) by the successful transmission probability $\xi$ to obtain the expected rate whose distribution for the working example is shown in Fig.~\ref{fig_collage_selecting_mcs}.

\begin{figure}[t!]
    \centering
    \includegraphics[width=0.5\textwidth]{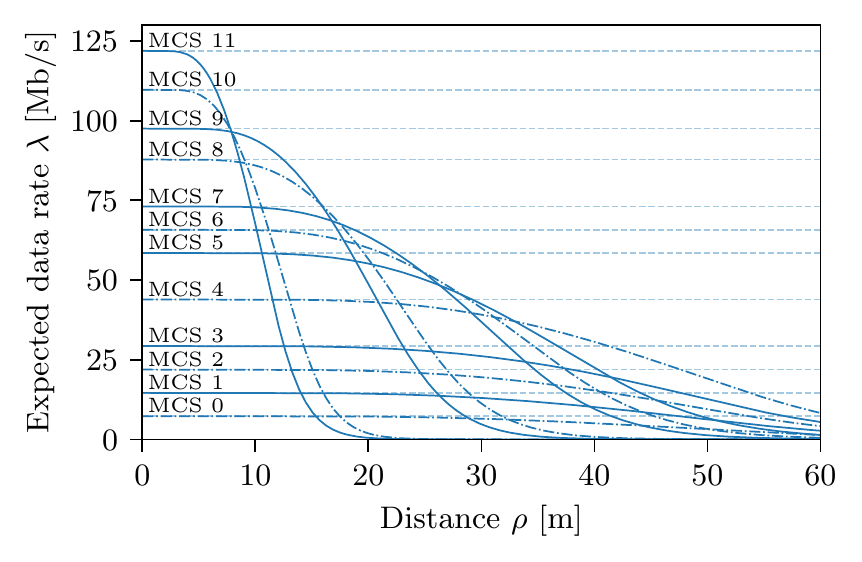}
    \caption{Relationship between SNR, mapped to distance according to \eqref{eq:dist_to_snr}, and the expected data rate for all single-stream IEEE 802.11ax MCS values. The curves were fitted to data obtained from ns-3 simulations for a log-distance path loss model and Nakagami multipath fading.}
    \label{fig:p_s}
\end{figure}

Each of these steps could be considered a trainable parametric model, so it introduces an error.
Therefore, we represent the intermediate values as random variables in the graphical model (Fig.~\ref{fig:graphical}).
Nevertheless, we observe that the errors are small and can be omitted in the first approximation. %
Since all transformations are bijective (Fig.~\ref{fig_collage}), so is their composition, as shown in Fig.~\ref{fig:p_s}. 
Thus, the approximation mentioned above allows us to directly map the distribution $p(\rho_t|\rho_{t<})$ to the rate distribution at each MCS.
The main benefit is to map the distance uncertainty $\rho$ to the uncertainty of the selected rate.
Since this transformation is highly non-linear, the variance of distance measurements shifts the expected rate.
This may cause a suboptimal MCS to be selected when operating close to the crossing point of two curves in Fig.~\ref{fig:p_s} (e.g., for $\rho\sim \SI{10}{m}$ three MCSs have similar rates).
The expectation in~\eqref{eq:opt} is computed after non-linearities as an average of multiple samples because, to the best of our knowledge, this non-linear transform has no known analytical result for expectation.

\begin{figure}
\centering
\subfloat[Stations are placed at a fixed distance from the AP]
{
    \includegraphics[width=0.8\columnwidth]{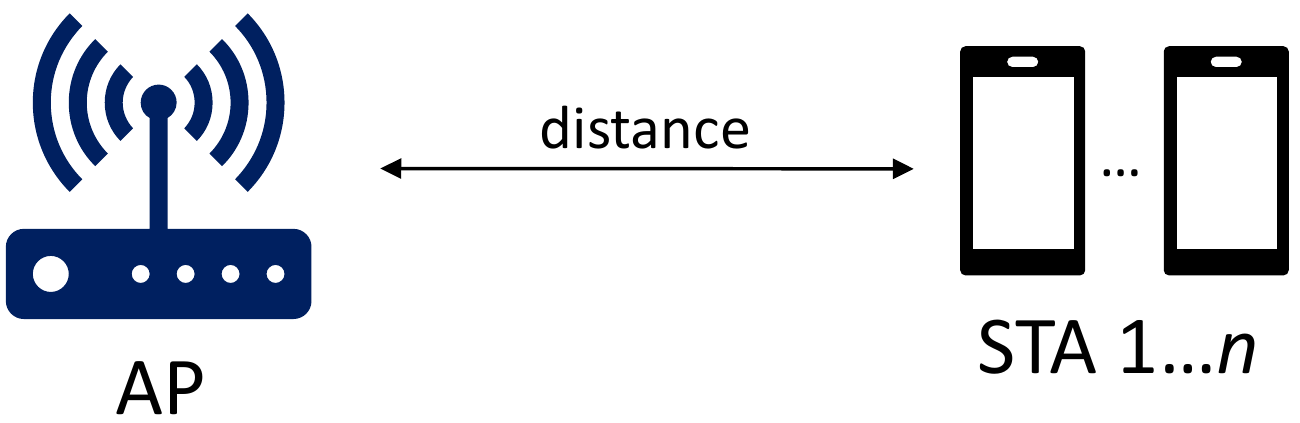}
    \label{fig:topo-fixed-distance}
}
\\
\subfloat[A single station moves away from the AP at a constant velocity.]
{
    \includegraphics[width=0.8\columnwidth]{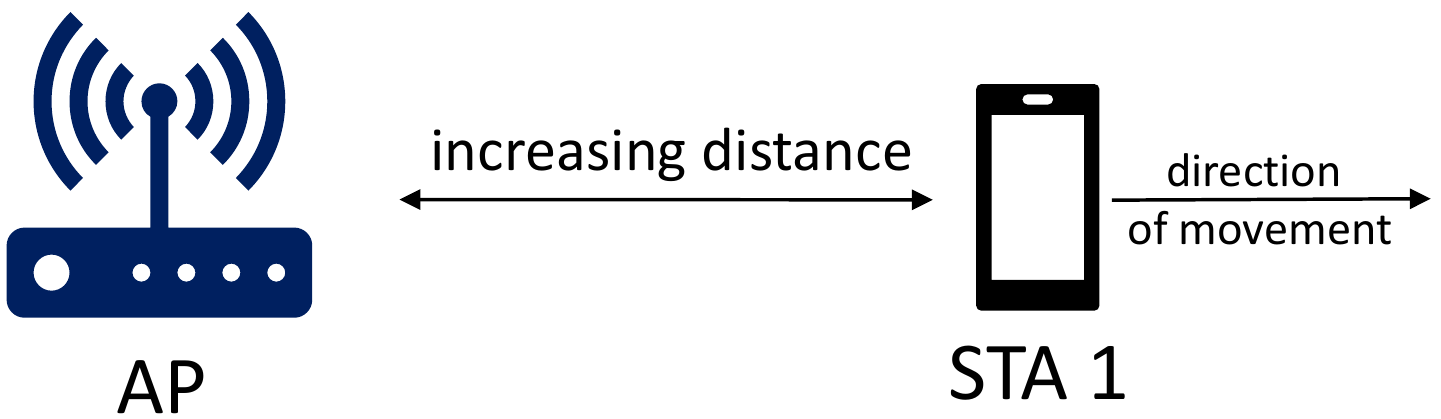}
    \label{fig:topo-single-station}
}
\\
\subfloat[Stations move according to the RWPM model and the AP is placed in the middle of a square area.]
{
    \includegraphics[width=0.8\columnwidth]{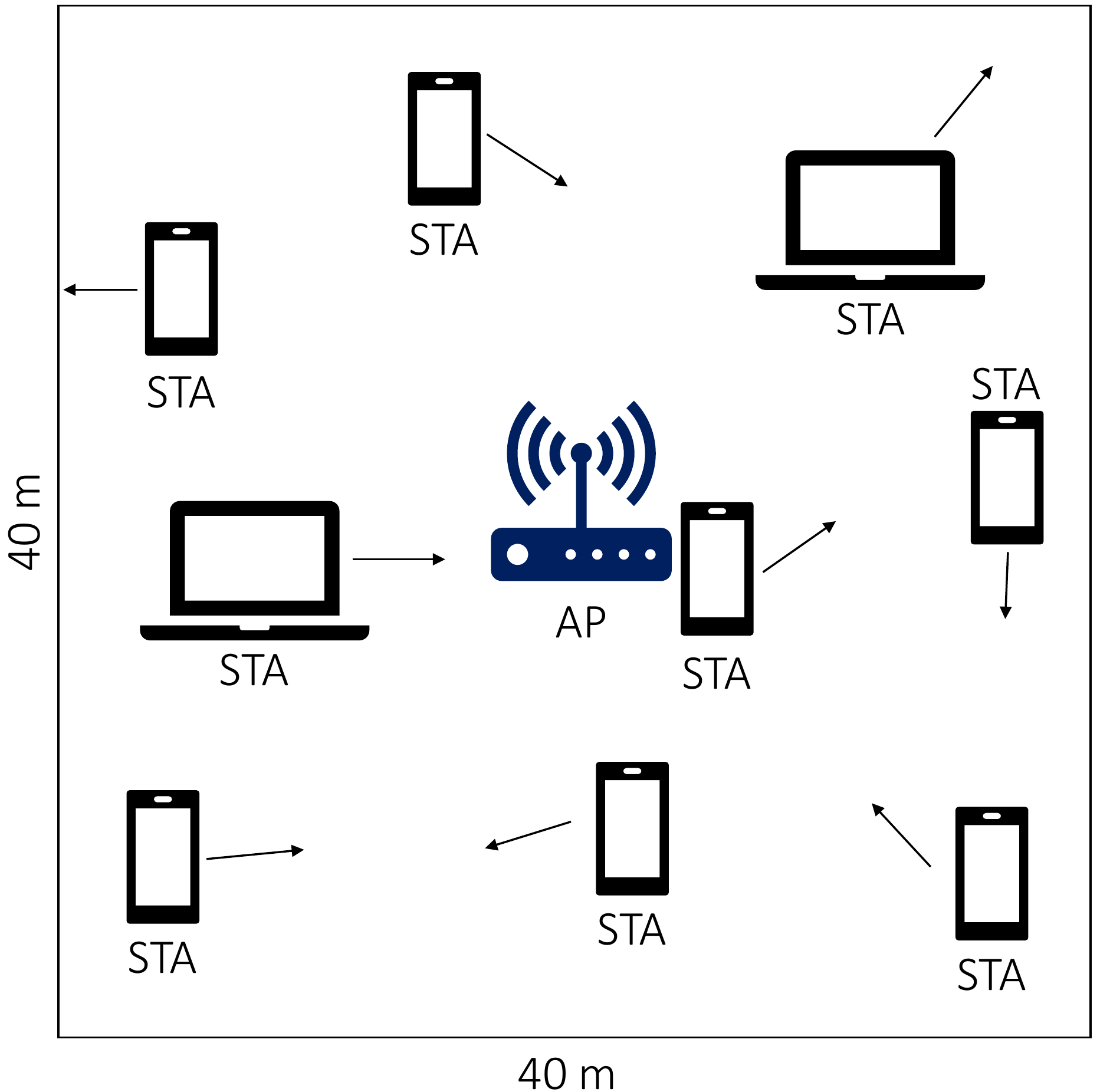}
    \label{fig:topo-random-distance}
}
\caption{Topologies of the (a) equal distance, (b) moving station, and (c) mobile station scenarios.}
\label{fig:topologies}
\end{figure}

\section{Performance Evaluation}
\label{sec:Scenarios}

We use the ns-3.36.1\footnote{\url{https://www.nsnam.org/}} network simulator to evaluate FTMRate.
Our proposal is implemented in Python, using the JAX library \cite{jax2018github},
and interfaces with ns-3 using ns3-ai \cite{yin2020ns3ai}.
We provide the complete code used in this study as open source\footnote{\url{https://github.com/ml4wifi-devs/ftmrate}}.

We evaluate FTMRate in its three inference variants: exponential smoothing, Kalman filter, and particle filter. 
As baseline rate selection protocols, we select the following:
\begin{itemize}
    \item Minstrel -- the default, non-ML rate selection method in Linux systems,
    \item Thompson sampling -- an
    ML-based rate selection method available in ns-3 \cite{krotov2020},
    \item oracle -- an artificial rate selection method that uses perfect destination knowledge 
    to calculate SNR according to the channel model of (\ref{eq:dist_to_snr}) and select MCS based on this knowledge. It does not consider multipath fading effects, so it is an approximation of the upper bound.
\end{itemize}

We consider both static and mobile scenarios with full-buffer stations transmitting to a single AP. 
We also assume no outside interference.
The general simulation settings are given in Table~\ref{tab_general_simulation_settings}, while the scenario-specific  settings are given in Table~\ref{tab_scenario_settings}.
To decrease the duration of the experiment, we use only 20~MHz channels (in the absence of overlapping networks, the results are qualitatively identical to using wider channel widths).
We present the 99\% confidence intervals as bands around the data points.

\begin{table}[t!]
\caption{General simulation parameter settings}
\label{tab_general_simulation_settings}
\centering
\begin{tabular}{@{}ll@{}}
\toprule
Parameter                  & Value                    \\ \midrule
Band                       & 5   GHz                  \\
PHY/MAC                    & IEEE   802.11ax          \\
Channel   width            & 20   MHz                 \\
Spatial   streams          & 1                        \\
Guard   interval           & 3200   ns                \\
Frame   aggregation        & A-MPDU aggregation                     \\
Loss   model               & log-distance w/ Nakagami fading \\
Path loss exponent $E$     & 3 \\
Reference SNR $\gamma_0$   & \SI{109.9906}{dB} \\
Reference path loss $L_0$   & \SI{46.6777}{dB} \\
Per-station   traffic load & Uplink,   125 Mb/s       \\
Packet   size              & 1500   B                 \\ \bottomrule
\end{tabular}
\end{table}

\begin{table*}[t!]
\caption{Simulation parameter settings for each scenario}
\label{tab_scenario_settings}
\centering
\begin{tabular}{@{}llll@{}}
\toprule
Scenario         & Equal distance                     & Moving station                & Mobile stations (RWPM)                            \\ \midrule
Repetitions      & 10                                 & 15                            & 40                              \\
Stations         & \{1, ..., 30\}           & 1                             & 10                              \\
Distance from AP & \{0, 20\} m                    & Increasing, constant velocity $\nu$ & Random walk (40 x 40 m)         \\
Velocity         & 0 m/s                              & \{1, 2\} m/s                  & $0-1.4$ m/s with $0-20$ s pause \\
Start position   & Fixed distance from AP             & 0 m                           & Random                          \\
Simulation time  & No. of stations $\times 10 + 50$ s & \{50, 25\} s                  & 1000 s                          \\ \bottomrule
\end{tabular}
\end{table*}

\begin{figure}[t!]
    \centering
    \includegraphics[width=\columnwidth]{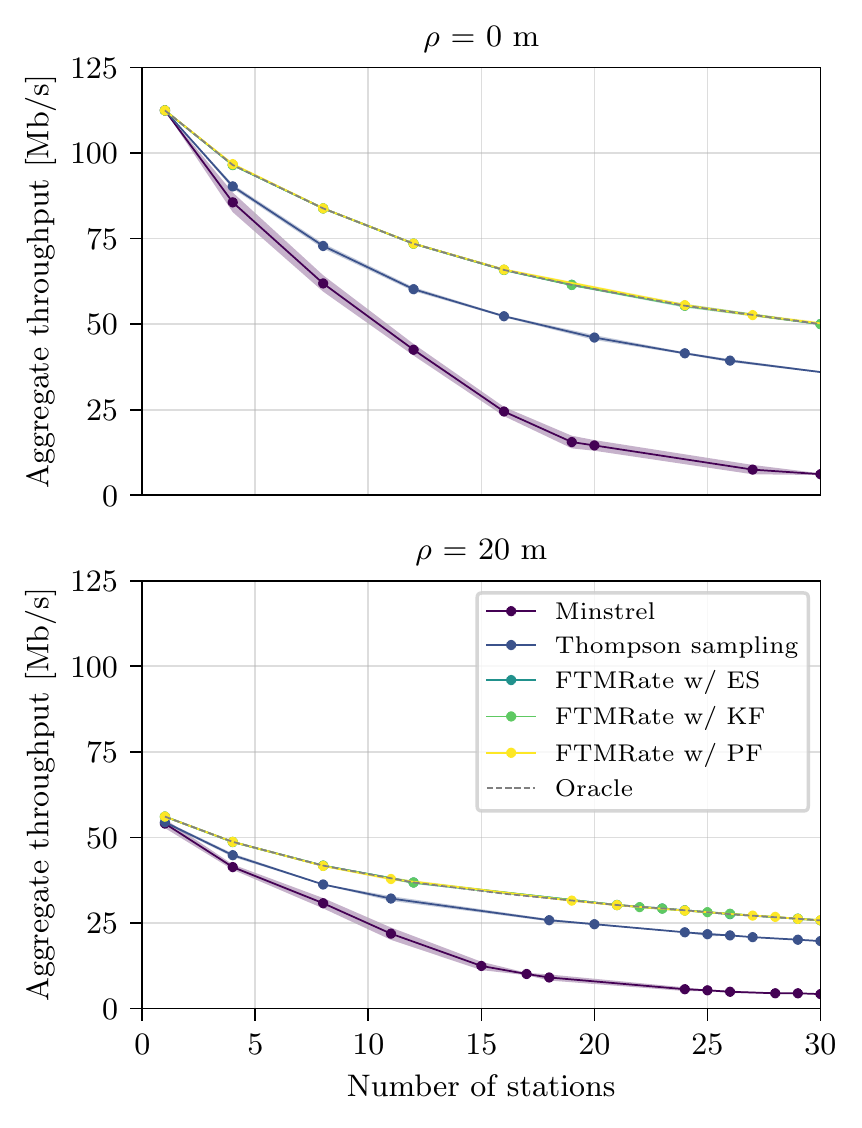}
    \caption{Aggregate network throughput in the equal distance scenario (Fig.~\ref{fig:topo-fixed-distance}): $\rho = \SI{0}{\meter}$ (top) and $\rho = \SI{20}{\meter}$ (bottom).}
    \label{fig:static}
\end{figure}

\begin{figure}[t!]
    \centering
    \includegraphics[width=\columnwidth]{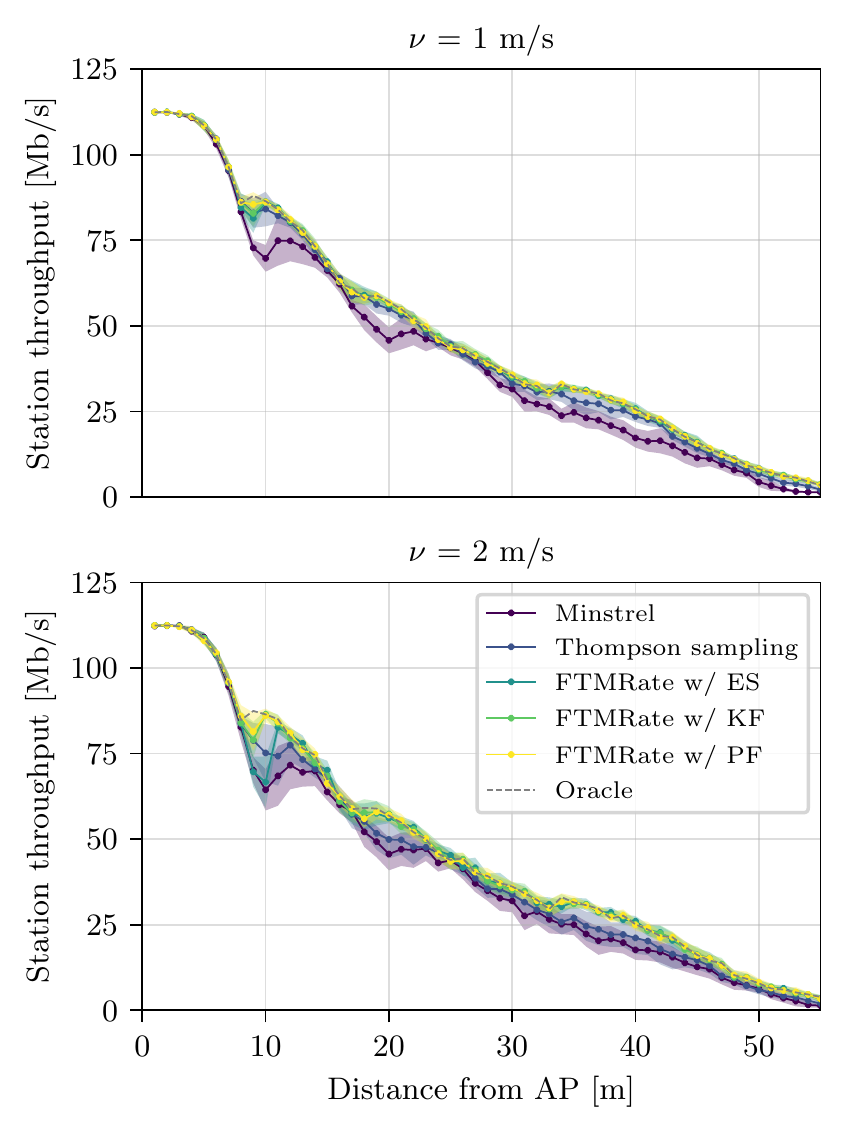}
    \caption{Throughput in the moving station scenario (Fig.~\ref{fig:topo-single-station}). A single station moves away from the AP at a constant velocity $\nu = \SI{1}{\meter/\second}$ (top) and $\nu = \SI{2}{\meter/\second}$ (bottom).}
    \label{fig:moving}
\end{figure}

\begin{figure}[t!]
    \centering
    \includegraphics[width=\columnwidth]{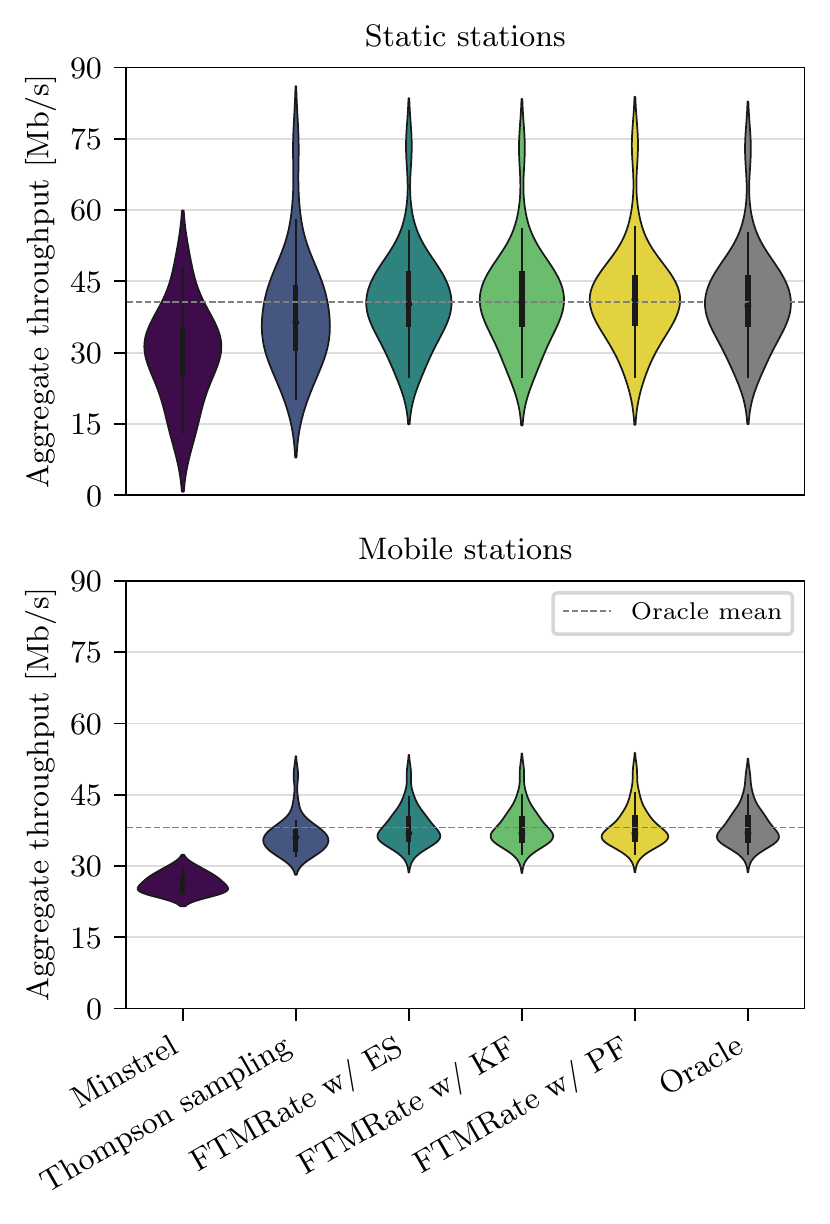}
    \caption{Aggregate network throughput with stations deployed randomly around the AP (Fig.~\ref{fig:topo-random-distance}): static stations ($\nu = \SI{0}{\meter/\second}$) with random distance (top) and mobile stations ($\nu = 0\dots\SI{1.4}{\meter/\second}$) in the RWPM scenario (bottom).}
    \label{fig:rwpm}
\end{figure}

\subsection{Equal Distance}
We begin with a scenario in which the stations are immobile and placed at a given distance from the AP, identical for each station (Fig.~\ref{fig:topo-fixed-distance}).
We analyze two distances $\rho = \SI{0}{\meter}$ and $\rho = \SI{20}{\meter}$.
The former setting is a theoretical case where the stations are co-located with the AP. Therefore, transmissions are under perfect channel conditions; however, collisions may still occur.
Therefore, stations should always select the highest MCS, i.e., do not mistake collisions for poor channel conditions.
In the latter setting, both collisions and channel errors occur (though without hidden stations).
At this distance, according to \eqref{eq:opt}, an MCS value of 7 is optimal.

The results confirm the well-known effect that the aggregate network throughput decreases with the number of transmitting stations (Fig.~\ref{fig:static}).
The performance of Minstrel and Thompson sampling suffers from the impact of collisions on the probability of successful transmissions: their throughput decreases faster than it would from the impact of collisions alone. Our approach is resilient to this problem because it does not consider whether transmissions are successful, but only considers FTM measurements.
Therefore, FTMRate, in all its inference variants, can reach the approximate upper bound (the oracle).
We conclude that FTMRate performs better in dense scenarios.

\subsection{Moving Station}
Next, we study a mobility scenario in which a single station moves away from the AP at a constant velocity $\nu$. 
This velocity is close to the human walking speed: $\SI{1}{}$ or~$\SI{2}{\meter/\second}$ (Fig.~\ref{fig:topo-single-station}). 
The station's throughput decreases as the distance between the station and AP increases (Fig.~\ref{fig:moving}). 
Most rate selection methods exhibit similar performance (only Minstrel is noticeably worse).
We conclude that the performance of FTMRate is no worse than that of competing solutions, even though our design goal is network density, not mobility.

\subsection{Static and Mobile Stations}
Finally, we study a scenario with random waypoint mobility (RWPM) of stations.
We place 10 stations at a random distance from the AP in a $\SI{40}{\meter} \times \SI{40}{\meter}$ square (Fig.~\ref{fig:topo-random-distance}).
Simulations are performed for the same station placement for all compared rate selection methods. In the first setting, static stations are configured (i.e., only random station placement around AP is introduced). In the second setting, mobile stations are configured (i.e., stations perform random walks, their velocity is randomly and uniformly selected from the range $\SI{0}{\meter/\second}$ to $\SI{1.4}{\meter/\second}$ and the pause time -- from $\SI{0}{\second}$ to $\SI{20}{\second}$). 

Fig.~\ref{fig:rwpm} presents the aggregate network throughput distributions (over all simulation runs) for each rate selection method. 
The wide distribution is caused by the high randomness of this scenario. 
In both static and mobile cases, the performance of Minstrel and Thompson sampling is affected by collisions, whereas the performance of the proposed solutions is close to the oracle.
Indeed, a Student's t-test reveals that, in the mobile case, only the difference between the means of the oracle compared to the means of Minstrel or TS is statistically significant (at a significance level of 0.05 with a p-value lower than 0.002).

\subsection{Discussion of Inference Variants}
The presented network performance results of FTMRate are quite similar for all three implemented inference variants.
Therefore, we proceed with a discussion of the advantages and disadvantages of using one specific approach among the three.
First, ES is extremely simple and lightweight but has several strict requirements: a constant sampling rate (ES lacks a proper probabilistic interpretation for continuous time), lack of probabilistic ouput, %
and setting the $\alpha$ and $\beta$ hyperparameters before use.
Next, KF is more sophisticated, but still rather lightweight. 
KF returns the result as a distribution that allows, e.g.,  adaptive measurements, and has an analytical solution for posterior inference \cite{murphy2012machine}.
Its underlying assumptions include a linear Markovian model of the system and a Gaussian distribution of error. 
Additionally, KF requires \textit{a priori} knowledge of sensor noise.
Finally, the third option, PF, can model any distribution of error (e.g., exponential gaussian) and also returns the result as a distribution.
However, it is computationally heavy (depending on the number of particles), and no analytical solutions exist (only approximations).
We plan to further investigate the performance differences between these inference variants.

\section{Conclusions}
\label{sec:conclusions}
We have presented FTMRate, a new data rate selection algorithm for IEEE 802.11 networks. 
FTMRate applies statistical learning to FTM distance measurements to first estimate the distance from the AP, then estimate the channel quality (SNR), and finally map SNR to MCS (and thus transmission rate).
The used approach makes FTMRate resilient to collisions and we have shown, in simulations of IEEE 802.11ax networks, that it can outperform existing approaches.
Our proposal, however, has certain limitations, including a signaling overhead and line-of-sight operation.

We note that by observing any random variable on the \textit{rate inference} path in Fig.~\ref{fig:graphical}, we can transform the problem into channel inference.
Therefore, as future work, we plan to investigate
methods to estimate channel parameters (including the path loss exponent) as well as optimize measurement frequency and handle cases where stations are in non-line-of-sight positions (e.g., behind a wall).
Furthermore, we want to study denser networks, include errors when estimating SNR, support dynamic power settings, and study fairness and energy consumption issues.
We believe that the signaling overhead could be meaningfully reduced, e.g., by appropriate station grouping for distance estimation, while mobility models could be improved, e.g., by moving to two dimensions and a non-linear model.


\end{document}